\documentstyle[twoside]{article}
\begin{document}
%%%%%%%%%%%%%%%%%%%%%%%%%%%%%%%%%%%%%%%%
%%%%%%%%%%%%%%%\def\baselinestretch{1.5}
\catcode`@=11
\def\marginnote#1{}
%%%%%%%%%%%%%%%%%%%%%%%%%%%%%%%%%%%%%%%%%
%
% definitions
\newcommand{\newc}{\newcommand}
%% macros to produce the symbols "less than or of order of"
%% and "greater than or of order of" %
\newc{\be}{\begin{equation}}
\newc{\ee}{\end{equation}}
\newc{\bea}{\begin{eqnarray}}
\newc{\eea}{\end{eqnarray}}
\newc{\gsim}{\lower.7ex\hbox{$\;\stackrel{\textstyle>}{\sim}\;$}}
\newc{\lsim}{\lower.7ex\hbox{$\;\stackrel{\textstyle<}{\sim}\;$}}
\newc{\thw}{\theta_W}
%%\newc\wd{\widetilde}
\newc{\ra}{\rightarrow}
\newc{\VEV}[1]{\langle #1 \rangle}
\newc{\hc}{{\it h.c.}}
\newc{\ie}{{\it i.e.}}
\newc{\etal}{{\it et al.}}
\newc{\eg}{{\it e.g.}}
\newc{\etc}{{\it etc.}}
\newc{\vrot}{v_{\rm rot}(r)}
\newc{\rhocrit}{\rho_{crit}}
\newc{\rhochi}{\rho_{\chi}}
\newc{\mpc}{~{\rm Mpc}}
\newc{\ev}{~{\rm eV}}     \newc{\kev}{~{\rm keV}}
\newc{\mev}{~{\rm MeV}}   \newc{\gev}{~{\rm GeV}}
\newc{\tev}{~{\rm TeV}}
\newc{\abund}{\Omega h^2_0}
\newc{\mx}{M_{GUT}}
\newc{\msusy}{M_{SUSY}}
\newc{\ewmgg}{SU(2)_L\times U(1)_Y}
\newc{\smgg}{SU(3)_c\times SU(2)_L\times U(1)_Y}
\newc{\mtwo}{M_2}
\newc{\mone}{M_1}
\newc{\tanb}{\tan\beta}
\newc{\mw}{m_W}   \newc{\mz}{m_Z}
\newc{\mchi}{m_{\chi}}
\newc{\hinot}{\widetilde H^0_2}
\newc{\hinob}{\widetilde H^0_1}
\newc{\bino}{\widetilde B^0}
\newc{\wino}{\widetilde W^0_3}
\newc{\hinos}{\widetilde H_S}       \newc{\hinoa}{\widetilde H_A}
\newc{\mcharone}{m_{\charone}}	\newc{\charone}{\chi_1^\pm}
\newc{\gluino}{\widetilde g}
\newc{\mgluino}{m_{\gluino}}
\newc{\photino}{\widetilde\gamma}
\newc{\flr}{ f_{L,R} }
\newc{\sfermion}{\widetilde f}
\newc{\msf}{m_{\sfermion}}
\newc{\snu}{\widetilde\nu}
\newc{\sel}{\widetilde e}
\newc{\msl}{m_{\widetilde l}}
\newc{\msq}{m_{\widetilde q}}
\newc{\msel}{m_{\sel}}
\newc{\mhalf}{m_{1/2}}
\newc{\mnot}{m_0}
\newc{\munot}{\mu_0}
\newc{\rgut}{r_{GUT}}
\newc{\mtop}{m_t}
\newc{\mbot}{m_b}
\newc{\hone}{H_1}
\newc{\htwo}{H_2}
\newc{\vev}{{\it v.e.v.}}
\newc{\vone}{v_1}   \newc{\vtwo}{v_2}
\newc{\hl}{h}   \newc{\hh}{H}   \newc{\ha}{A}
\newc{\mhl}{m_\hl}   \newc{\mhh}{m_\hh}   \newc{\ma}{m_A}
\newc{\ch}{C}   \newc{\chpm}{C^{\pm}}   \newc{\chmp}{C^{\mp}}
\newc{\mch}{m_\ch}   \newc{\mchpm}{m_\chpm}   \newc{\mchmp}{m_\chmp}
\def\NPB#1#2#3{Nucl. Phys. B {\bf#1} (19#2) #3}
\def\PLB#1#2#3{Phys. Lett. B {\bf#1} (19#2) #3}
\def\PLBold#1#2#3{Phys. Lett. {\bf#1B} (19#2) #3}
\def\PRD#1#2#3{Phys. Rev. D {\bf#1} (19#2) #3}
\def\PRL#1#2#3{Phys. Rev. Lett. {\bf#1} (19#2) #3}
\def\PRT#1#2#3{Phys. Rep. {\bf#1} C (19#2) #3}
\def\ARAA#1#2#3{Ann. Rev. Astron. Astrophys. {\bf#1} (19#2) #3}
\def\ARNP#1#2#3{Ann. Rev. Nucl. Part. Sci. {\bf#1} (19#2) #3}
\def\MODA#1#2#3{Mod. Phys. Lett. A {\bf#1} (19#2) #3}
\def\APJ#1#2#3{Ap. J. {\bf#1} (19#2) #3}
%%%%%%%%%%%%% end of defs %%%%%%%%%%%%%%%
\begin{titlepage}
\begin{center}
\hfill    UM-TH-93-06 \\
\hfill
hep-ph/9302259\\
\hfill    February 1993\\
\vskip 0.75in
{\large \bf
SUPERSYMMETRIC DARK MATTER -- A REVIEW
\footnote{\em Based on invited talks at the XXXII Cracow School on
Theoretical
Physics, Zakopane, Poland, June~2 - 12, 1992; the
23rd Workshop ``Properties of SUSY Particles'', Erice, Italy,
September~28 - October~4, 1992; and (partly) the 1993 Aspen Winter
Conference on High Energy Physics, January~11 - 16, 1993;
to appear in the Proceedings of the Erice Workshop.}
}
\vskip .4in
{\large LESZEK ROSZKOWSKI}
\vskip .1in
{\em leszek@leszek.physics.lsa.umich.edu\\
     Randall Physics Laboratory,\\
     University of Michigan,\\
     Ann Arbor, MI 48109-1129, USA}
\end{center}
\vskip .2in
\begin{abstract}
\noindent
I address the question of whether supersymmetry provides a viable
candidate for the dark matter in the Universe.
I review the properties of the lightest
neutralino as a candidate for solving
the dark matter problem.
I discuss the neutralino's phenomenological and cosmological
properties, and constraints from present and future experiments.
In the minimal supersymmetric model, the neutralino mass has been
experimentally excluded below some $20$ GeV, and is not expected to
be
significantly larger than about $150$ GeV.
I identify a gaugino-like neutralino as the most natural dark matter
candidate
for a plausible range of parameters. The requirement that the
lightest neutralino be the dominant matter component in the flat
Universe provides non-trivial restrictions on other parameters of
the model, in particular on the masses of the sfermions.
Next, I study the consequences of adopting further grand unification
assumptions.
In both scenarios I find sfermion masses most likely beyond the
reach of LEP 200 and the Tevatron but well within the discovery
potential of the SSC and the LHC.
I also comment on  the effects of relaxing grand unification
assumptions.
Finally, I briefly outline prospects for the neutralino dark matter
searches.
\end{abstract}

\end{titlepage}
\setcounter{footnote}{0}
\setcounter{page}{2}
\setcounter{section}{0}
\setcounter{subsection}{0}
\tableofcontents
\newpage
% BODY

\section{Introduction}

Experimentally supersymmetry (SUSY) still remains an open
question,
many theorists now express their growing confidence that it
is indeed realized in Nature.
Originally applied to high energy theories to
solve the naturalness problem, supersymmetry has actually proven
very successful in several other aspects of particle
physics~\cite{susyrev,hk}.
Indeed,
even minimal supersymmetric grand unified model predicts the right
ratio of
gauge couplings~\cite{susyunif}, as recently confirmed by
LEP~\cite{amaldi};
provides a mechanism for dynamical electroweak gauge symmetry
breaking~\cite{gsymbreak}; generically predicts the proton decay
beyond
the present experimental reach~\cite{protondecay}; and provides a
nice
connection with theories valid
at the Planck mass scale, like supergravity and superstrings.
At the same time, present experimental constraints on even minimal
version of
supersymmetry, both
via direct accelerator searches and through supersymmetric loop
contributions,
are so far relatively mild~\cite{susyexptalrev}. It has been also
recognized
for some time~\cite{ehnos} that supersymmetry also provides a nice
solution
to the
dark matter (DM) hypothesis which states that perhaps more than
$90\%$ of matter
in the Universe is non-shining (see below and, \eg,
Refs.~\cite{dmrev,kt}).

In this report, I will review the present status of the lightest
neutralino,
assumed to be the lightest supersymmetric particle (LSP),
as a DM candidate.
The literature on the subject is already very
extensive and I will have to make some crude selections. I will first
summarize the main results in the framework of minimal supersymmetry.
Next I will emphasize more recent work that has been
done in the context of the minimal supergravity model.

In some contrast to more  astrophysically and cosmologically oriented
reviews on supersymmetric dark matter
(see, \eg, Ref.~\cite{turnerinsm}),
in considering the LSP as dark
matter I will make a fuller use of other, experimental
and theoretical, constraints of supersymmetry, including various
relationships between SUSY parameters.

I will start by swift reviews of both the DM problem and
supersymmetry.
In chapter~\ref{mssm} I will briefly
introduce the minimal supersymmetric model, and will define the
lightest
supersymmetric particle. I will summarize its properties and
experimental
constraints. In chapter~\ref{lspdm} I will discuss various
cosmological  properties of the neutralino and the associated
constraints on other supersymmetric parameters, in particular on the
masses of the sfermions. Next in chapter~\ref{gutsect} I will review
the implications
that the requirement that the LSP be the DM in the Universe has on
the highly constrained grand unification scenario.
I will
close this chapter with a brief discussion of the implications of
relaxing the grand unification assumptions.
In chapter~\ref{dmsearches} I briefly outline the present status of
direct and
indirect searches for dark matter. Some general comments and a brief
summary will close this review.

\subsection{The Dark Matter Problem}\label{dmproblem}

There is increasing astrophysical evidence that most of the matter in
the Universe is dark, \ie, does not emit nor absorb electromagnetic
radiation (at least at the detectable level).
This evidence, while
only circumstantial, comes
from vastly different cosmological scales, ranging from galactic
scales of several kiloparsecs
(1pc$=3.26$~light-year~$=3.1\times10^{16}$m) to clusters of galaxies
(several megaparsecs), and up to global scales of hundreds of Mpc's.
There exist several extensive reviews on the
subject~\cite{dmrev,kt,turnerinsm},
and I will quote but a few examples.
The most well-known evidence comes from observations of rotational
velocities $\vrot$ of spiral galaxies. If the shining matter
were the total matter of the galaxies then one would expect
$\vrot\propto1/\sqrt{r}$.
Instead, one observes $\vrot\approx {\rm const}$ which implies that
the total mass of a galaxy within radius $r$,
$M_{\rm DM}(r)$ grows like $M_{\rm DM}(r)\propto r$. One then
concludes
that there  exist extensive galactic halos consisting of DM.
Estimates show that $M_{\rm DM}/M_{\rm
vis}\gsim 3-10$, and perhaps even more. There is somewhat
more uncertain
evidence on the scale of 10 to 50\mpc\ (clusters of galaxies) for
even
more DM. Finally, some first evidence has been provided by several
groups~\cite{turnerinsm}
(like POTENT~\cite{potent})
that over very large scales (hundreds of megaparsecs) the total mass
density $\rho$ approaches the critical density $\rhocrit$.
The critical density
$\rhocrit\equiv{3H_0^2}/{8\pi G}=1.9\times10^{-29} (h_0^2) g/cm^3$
corresponds to the flat Universe, and $h_0$ is the present value of
the
Hubble parameter $H_0$ in units 100~km/s/Mpc. The value
$\Omega\equiv\rho/\rhocrit=1$ is strongly preferred by theory since
it is predicted by the models of
cosmic inflation and is the only stable value for
Friedmann-Robertson-Walker
cosmologies. Larger values of $\Omega$ are also strongly supported by
present models of
primordial barygenesis and
most models of large structure formation.

The visible matter in the Universe accounts for less than $1\%$ of
the
critical density.
Primordial nucleosynthesis constrains the allowed range of baryonic
matter in the Universe to the range $0.02<\Omega_b<0.11$~\cite{bbn}
(and more recently~\cite{bbn} $\Omega_b\approx 0.05$).
This, along with estimates given above, implies that: {\em (i)} most
baryonic matter in the Universe is invisible to us, and
{\em (ii)} already
in halos of galaxies one might need
a substantial amount of non-baryonic DM.
This is the galactic DM problem.

If $\Omega=1$ then most (about 95\%) of the matter in
the Universe is probably non-baryonic and dark.
This is the global DM problem.
Current estimates show that, for $\Omega=1$,
$0.5\lsim h_0\lsim0.7$ (the upper bound coming from
assuming the age of the Universe above 10 bln years),
in which case one expects the DM abundance in the range
\be
\label{abundrange}
0.25\lsim\abund\lsim0.5.
\ee
More conservatively (\ie, without assuming $\Omega=1$, but still
assuming the age of the Universe above
10 bln years), one excludes $\abund>1$.
On the other hand, the density of
DM should be at least $\abund\gsim0.025$ in order to provide minimum
required DM at least in galactic halos.

Finally, one should keep in mind that the uncertainties involved in
measuring both $\Omega$
and $h_0$ are large and the bounds on $\abund$ quoted above should
not be strictly interpreted. (In fact, various authors adopt somewhat
different
criteria -- this doesn't cause the conclusions to be dramatically
different). On the other hand, the evidence that DM is
abundant at various length scales in the Universe is now so
convincing that it would be very surprising if it did not prove true.

One more remark about the specific nature of DM should be made in
closing
this section. Before the recent COBE discovery of the cosmic
microwave
background anisotropies~\cite{cobe}
(for an elementary review, see, \eg, Ref.~\cite{mycobe}), models
of large structure formation widely favored so-called cold DM, rather
than hot DM (\ie, non-relativistic and relativistic, respectively,
see section~\ref{lspdm}). The all-cold-DM case, while still very
successful, has become increasingly at odds with the measured angular
correlation function of galaxies and with their pair-wise velocity
dispersions. In the aftermath of the COBE discovery, some people have
argued for a mixed DM scenario with roughly 60-70\% of cold DM and
about
30\%
of hot DM, like neutrinos with mass in the range of a few eV. (Others
favor the non-zero cosmological constant as a main contributor to
$\Omega=1$.) These issues are far from being fully clarified but they
won't have any truly dramatic effect on the results presented here.
Conservatively, all one really assumes by adopting the
range~(\ref{abundrange}) is that the LSP contributes significantly to
the total matter density in the flat Universe.

\subsection{Supersymmetry}\label{susy}

Supersymmetry~\cite{susyrev,hk} has become popular, in part,
because it solves one of the most
serious problems of the Standard Model of electroweak and strong
interactions, namely the naturalness and the fine-tuning problems,
as advocated early, among others, by Veltman~\cite{tini}.

Supersymmetry must be broken at some effective scale $\msusy$
since the experimental bounds on the masses
of the scalar (\ie, spin zero boson)
partners of ordinary fermions are experimentally known
to be higher then the masses of the fermions themselves.
The commonly accepted way of breaking global supersymmetry is to
introduce terms which explicitly break it but in such a way that no
quadratic divergences are re-generated. Such terms are often called
`soft' and are expected to be of the order of $\msusy$
which sets the order of magnitude for the Higgs mass, and
therefore they should
not significantly exceed the Fermi mass scale. Thus in this weakly
(or
softly) broken supersymmetry scenario, the masses of the fermions and
gauge bosons are given in terms of the Higgs boson's vacuum
expectation
value, while the masses of the scalars are set roughly by $\msusy$.

In the most commonly accepted approach one starts with
a grand unified theory with unbroken SUSY above some scale
$\mx\approx10^{16}\gev$. At that scale, the GUT gauge symmetry breaks
down to the Standard Model gauge symmetry, but supersymmetry remains
unbroken until $\msusy\sim{\cal O}(1\tev)$. Below that scale, the
breakdown of
SUSY due to the soft terms
induces a non-zero  Higgs vacuum expectation value, and
thus triggers the spontaneous
breakdown of the electroweak gauge symmetry
$\ewmgg$ down to the $U(1)$ of electromagnetism.

The terms which break global supersymmetry explicitly but softly
arise naturally when a supersymmetric GUT is coupled to supergravity.
Supergravity is a theory of local supersymmetry which
provides a  framework for incorporating gravity into
a unified theory of all four fundamental interactions. It also
allows for a mechanism of global SUSY breaking. I will come back
to this issue in section~\ref{gutass} where I discuss the dark matter
problem in the context of minimal supergravity.

The program outlined above is realized even in the simplest
phenomenologically viable supersymmetric model, known as the minimal
supersymmetric model, or Minimal
Supersymmetric Standard Model.

\section{Minimal Supersymmetric Standard Model}\label{mssm}

\subsection{Generalities}\label{general}

The simplest supersymmetric theory of phenomenological (and
cosmological) interest is the Minimal Supersymmetric Standard Model
(MSSM) (for a review, see, \eg, Ref.~\cite{susyrev,hk,gh}).
In the MSSM, one adds to all the fields of the Standard Model
their supersymmetric partners to form supermultiplets.
The supersymmetric part
of the
Lagrangian results from the following superpotential
\be
\label{susylagr}
W=\sum_{generations} (h_U Q_L U_L^c\htwo
+ h_D Q_L D_L^c\hone + h_L L E_L^c\hone) -\mu\hone\htwo.
\ee
(I use here
the same notation for
both ordinary fields and their chiral superfields.) Other terms,
which could break B and/or L number conservation, are
absent once a discrete symmetry, the so-called $R$-parity, is
assumed.
In Eq.~(\ref{susylagr})
$h_{U,D,L}$ are (matrix) Yukawa couplings for the
up-type quarks, down-type quarks, and charged leptons, respectively,
and generation indices have been suppressed. The Higgs mass
parameter $\mu$ is assumed to be of the order of $\mz$.
SUSY requires two Higgs doublets: $\hone=(\hone^0,\hone^-)$ and
$\htwo=(\htwo^+,\htwo^0)$. Their neutral components $\hone^0$ and
$\htwo^0$ acquire \vev s $\vone$ and $\vtwo$ which give masses to
down
and up-type fermions, respectively.
The Higgs mass spectrum consists of two physical scalar fields $\hl$
and
$\hh$, one pseudoscalar $\ha$, and a pair of charged Higgs bosons
$\chpm$.
At the tree level the Higgs sector is fully described
in terms of just two parameters, which I take to be $\ma$ and
$\tanb\equiv\vtwo/\vone$.
The expected range of values for $\tanb$ lies between 1 and
$m_t/m_b$.
Expressions for the Higgs masses can be
found, \eg, in Ref.~\cite{gh}.
Here I only quote the well-known relations:
$\mhl\leq\mz|\cos2\beta|<\mz<\mhh$, $\mhl<\ma<\mhh$, and $\mch>\mw$.
Radiative corrections to the Higgs masses have recently been shown to
be potentially significant. I will discuss their implications later.

\subsection{Lightest Supersymmetric Particle}\label{lsp}

In the MSSM the four neutralinos $\chi^0_i$ ($i=1,...,4$)
are the physical (mass) superpositions of
two fermionic partners of the neutral Higgs bosons, called
higgsinos $\hinob$ and $\hinot$,
and of the two neutral gauge bosons, called
gauginos $\bino$ (bino) and $\wino$ (wino).
They are Majorana fermions which means that they are invariant under
charge conjugation.
The neutralino mass matrix is given by~\cite{ehnos,gh}
\vskip 0.05in
\be
%%\[
\left(
\begin{array}{cccc}
\mtwo & 0 & \mz\cos\thw\cos\beta & -\mz\cos\thw\sin\beta\\
0 & \mone & -\mz\sin\thw\cos\beta & \mz\sin\thw\sin\beta\\
\mz\cos\thw\cos\beta&-\mz\sin\thw\cos\beta&0&-\mu\\
-\mz\cos\thw\sin\beta&\mz\sin\thw\sin\beta&-\mu&0
\end{array}
\right).
%%\]
\label{massmat}
\ee
\vskip 0.05in

The lightest neutralino
\be
\chi\equiv
\chi_1^0=N_{11}\widetilde W^3+N_{12}\widetilde B+N_{13}\widetilde
 H_1^0+N_{14}\widetilde H_2^0
\label{chilsp}
\ee
will be henceforth assumed to be the lightest supersymmetric particle
(LSP).
This assumption is a plausible but arbitrary one in the MSSM but is
naturally realized in the low-energy limit of supersymmetric grand
unified theories as I will discuss in section~\ref{gutass}.
Due to the assumed $R$-parity, which assigns a value $R=-1$ to
sparticles, and $R=+1$ to ordinary particles,
the LSP is absolutely stable: it
cannot decay to anything lighter. It can, however, still annihilate
with
another sparticle (in particular with itself) into ordinary matter.

The neutralino parameter
space is described in terms of four quantities: $\tanb\equiv
v_2/v_1$,
the Higgs/higgsino mass parameter $\mu$, and the two gaugino mass
parameters
$\mone$ and $\mtwo$ of the $\bino$ and $\wino$ fields,
respectively~\cite{hk,gh,ehnos}.

Usually, one assumes that all gaugino masses are equal at
a grand unification theory (GUT) scale (see section~\ref{gutass}),
from which it follows that
\be
\mone= \alpha_1/\alpha_2\mtwo\approx0.5\mtwo,
\label{gutone}
\ee
as well as
\be
\mtwo=\alpha_2/\alpha_s\mgluino\approx 0.3\mgluino,
\label{guttwo}
\ee
where $\alpha_{1,2,s}$ are the gauge strengths of the groups
$U(1)_Y, SU(2)_L$ and $SU(3)_c$, respectively, and
$\mgluino$ is the mass of the
gluino $\gluino$, the fermionic partner of the gluon.

Charginos are the charged counter-partners of neutralinos.
Their masses are given by~\cite{ehnos,gh}
\begin{eqnarray}
m^2_{\chi^\pm_1,\chi^\pm_2}&=&
{1\over 2}\Bigl[\mtwo^2+{\mu}^2+2\mw^2\\
& &\mp\sqrt{(\mtwo^2-{\mu}^2)^2+4\mw^4\cos^2{2\beta}
+4\mw^2(\mtwo^2+{\mu}^2+2\mtwo\mu
\sin2\beta)}\Bigr],\nonumber
\label{charmasseq}
\end{eqnarray}
and, assuming the relation (\ref{gutone}), are described in terms of
the
same parameters as the neutralinos.

\subsection{Parameter Space of the LSP}
\label{thexp}

It is first worth recalling the phenomenological structure
of the neutralino parameter space. It is convenient to display
the mass and gaugino/higgsino composition contours in the plane
$(\mu,\mtwo)$ for discrete values of $\tanb$.
This is shown in
Fig.~1 for a typical value of $\tanb=2$.
For $|\mu|\gg\mtwo$, $\mchi\approx\mone\approx 0.5\mtwo$, and
the LSP is an almost pure gaugino (and mostly a bino $\bino$).
For $\mtwo\gg|\mu|$, $\mchi\approx|\mu|$, and the LSP
is a nearly pure higgsino. More specifically, it is $\hinos$ for
$\mu<0$ and $\hinoa$ for
$\mu>0$, where $\widetilde H_{S,A}\equiv [\pm\widetilde H_1
+\widetilde H_2]/\sqrt2$.
In the intermediate (`mixed') region the LSP
consists of comparable fractions of both gauginos and higgsinos.
To be quantitative, I adopt gaugino purity, defined
as $p_{gaugino}=Z_{11}^2 + Z_{12}^2$~\cite{chiasdm,nogut,chiatlep2}
of, say,
$90\%$ to distinguish
higgsino, gaugino, and `mixed' regions.
(For simplicity, below higgsino-like LSPs will often be called just
higgsinos, and analogously for gauginos.)
This distinction is important because cosmological
properties of higgsinos and gauginos are significantly different!

\subsection{Present and Future Experimental Constraints on the LSP}
\label{exptallims}

Large fractions of the $(\mu,\mtwo)$ plane have been excluded by LEP
(see, \eg, Refs.~\cite{nogut,dn}):
by direct searches for charginos ($\mcharone>46\gev$) and
neutralinos, and indirectly, from
their contributions to the $Z$ line shape at LEP.
If one assumes equal gaugino masses at the GUT scale then additional
regions are also excluded by
the unsuccessful CDF searches for the gluino.
An early preliminary bound of about $150\gev$ (with simplifying
assumptions~\cite{cdfgluino-prelim} excludes $\mtwo\lsim 44.7\gev$
and leads
to a lower bound~\cite{chimasslimit}
\be
\label{lowerlimonchi}
\mchi\gsim20\gev
\ee
on the mass of the lightest neutralino.
The bound Eq.~(\ref{lowerlimonchi}) also survives~\cite{chiatlep2}
possibly large corrections to the Higgs masses
due to the large top-quark mass. This is because larger statistics
accumulated during more recent runs of LEP has allowed
significantly larger regions of the $(\mu,\mtwo)$ plane to be
excluded also for values of $\tanb$ as small as one.

A more careful analysis
of the gluino bound which takes into account its possible decays into
heavier states which next cascade-decay into the LSP leads to
a somewhat weaker bound of about $135\gev$~\cite{tata},
or $90\gev$~\cite{cdfgluino},
which in turn scales the bound Eq.~(\ref{lowerlimonchi})
down to about $18\gev$ and $12\gev$, respectively.

Recent experimental results also strongly
constrain the LSP's properties.
First, since the neutralinos couple to the $Z$ through their higgsino
components only, it is not surprising that higgsino-like $\chi$'s
below
about $45\gev$ have been excluded by LEP.
Gaugino-like LSPs have only been indirectly constrained by chargino
and gluino
searches, with the result given in the bound~(\ref{lowerlimonchi}).
Moreover, the regions where the LSP is photino-like
(the photino
$\photino\equiv
\sin\theta_w\widetilde W^3+\cos\theta_w\widetilde B$
is never an exact mass eigenstate!) have been
excluded above the $96\%$ purity level~\cite{nogut,chiatlep2}
as can be seen from fig.~1.

Significant progress in exploring the neutralino parameter space is
expected to be made after LEP 200 starts operating in early
1995~\cite{chiatlep2,ghsnowmass}.
The strongest constraint will come from pushing the chargino searches
up to about
$80\gev$, which in turn will place indirect lower bounds of about
$40\gev$ on gaugino-like $\chi$'s and of about $80\gev$ on
higgsino-like $\chi$'s. This will have some cosmological
consequences~\cite{chiatlep2}
as I will discuss later.

Dark matter searches are also capable of exploring the neutralino
parameter space, if one assumes the LSP to be dominant component of
the galactic halo. I will discuss this in section~\ref{dmsearches}.

\subsection{Theoretical and Cosmological Upper Bounds on the LSP
Mass}\label{thcosmlims}

While the LSP mass is constrained by high-energy experiments from
below, one would also like to be able to restrict it from above.

Theoretically, the expectation that the SUSY breaking scale
(and hence also $\mgluino$) should not significantly exceed
the $1\tev$  scale in order to avoid the hierarchy problem leads, via
the GUT relation Eq.~(\ref{gutone}),
to a rough {\it upper} bound~\cite{chiasdm}
$\mchi\lsim 150\gev$, see fig.~1.
This upper limit
is only indicative (and it scales linearly with $\mgluino$), but sets
the overall scale for expected LSP masses.
It also coincides with a similar bound of about $110\gev$
(for $m_t>90\gev$) resulting from applying the naturalness
criterion~\cite{bg}.
(I will have more to say
about this in the context of minimal supergravity.)
This indicative upper bound also implies~\cite{chiasdm}
that {\em higgsino-like $\chi$'s are strongly disfavored},
since they correspond
to uncomfortably large gluino masses.
Experimental constraints and theoretical criteria (naturalness) then
point us towards the gaugino-like and `mixed' regions.

Interestingly,
one can derive also cosmological upper bounds on $\mchi$. This was
originally done by Olive, \etal~\cite{os}, and Griest,
\etal~\cite{gkt}
in the case of higgsino-like LSP of roughly
$\mchi\lsim2.8\tev$ and gaugino-like LSP of about
$\mchi\lsim550\gev$. These bounds correspond to very large
gluino masses of
about $20\tev$ and $3\tev$, respectively. (In the higgsino case the
co-annihilation of the LSP with charginos and other neutralinos in
fact
greatly {\it weakens} the bound, as I will discuss later.)
More recently, Drees and
Nojiri~\cite{dn} have sharpened the bound for the bino case to
$\mchi<350\gev$ in the context of
minimal supergravity. But they also estimated that, in special cases,
(enhanced pole annihilation) in the `mixed' region the constraint
$\abund<1$ allows for the LSPs as heavy as $100\tev$ (!), the range
not necessarily fitting the concept of low-energy supersymmetry.
One should also remember that in deriving such cosmological bounds
one assumes the the LSP is a {\em pure} higgsino or bino state which
in practice is never the case. In particular, in the grand-unified
scenario, where all the masses are inter-related, {\em no}
well-defined cosmological
bound has been found~\cite{dick}.
In brief, it is worth remembering that in some cases one can derive
cosmological upper bounds on $\mchi$ but they lie well beyond
theoretical expectations.

\section{Neutralino LSP as Dark Matter}\label{lspdm}

Particle candidates for dark matter can be divided
into two broad classes: hot and cold DM~\cite{kt,turnerinsm}.
The first were
relativistic at the time of `freeze-out' (see below), while the
latter
were non-relativistic. Light massive neutrinos, with the mass of
about
10\ev, are the well-known
(and well-motivated) candidates for hot DM. In
the cold DM class the leading candidates are the
neutralino LSP~\cite{ehnos}
in the mass range of several GeV, and the axion.
LEP has excluded several other candidates for DM,
like heavy neutrinos or their supersymmetric partners, sneutrinos, by
placing a lower bound of about $\mz/2$ on their mass.
This implies
that their contribution to the total mass of the
Universe is uninterestingly small ($\abund\lsim10^{-3}$).
Other experiments have severely narrowed the allowed mass range of
the axion~\cite{axion}.
In contrast,
the lightest neutralino $\chi$ has all the desired properties of
a natural candidate for both the
lightest supersymmetric particle (LSP) and the dark matter in
the Universe~\cite{ehnos}.
It is neutral, weakly-interacting, and stable (assuming R-parity).
In addition, it typically gives the relic
abundance in the Universe $\abund$
within an order of magnitude from unity.
In recent years important constraints on the
neutralino properties have been derived, and  various conditions
that need to be satisfied for the LSP to
be a solution to the DM problem for natural ranges of parameters have
been pointed out.
They will be summarized in this chapter. First I will briefly
outline the procedure of computing the relic density of neutralinos
and, in general, any relic species.

\subsection{Computing the Neutralino Relic Abundance}\label{compute}

In the very early Universe ($t\lsim10^{-12}s$), all the species,
including
the neutralinos, were in thermal equilibrium with
photons~\cite{kt,turnerinsm}.
At that stage the neutralinos could reduce their number
via pair annihilation $\chi\chi\rightarrow
{\rm ordinary\  matter}$.
As the Universe was expanding and cooling, the density of the LSPs
was
decreasing, and it was becoming
increasingly harder for them to find partners
to annihilate. At some temperature, called the `freeze-out'
temperature $T_f$, the number of the LSPs in the Universe became
effectively
constant, and approximately equal to their number today. In other
words, the LSP relic density per co-moving volume became
approximately constant.

The relic abundance $\abund$ of the neutralinos, or any other relic
species,
can be computed by solving the Boltzmann (rate) equation
\be
{{d n_\chi}\over{dt}} = -3H n_\chi - \VEV{\sigma v_{rel}}
\left[n_\chi^2-(n_\chi^{eq})^2\right],
\label{boltzmanneq}
\ee
where $n_\chi$ is the actual number density of the LSPs, and is
related
to $\abund$ by $\abund=(\mchi n_\chi h_0^2)/\rhocrit$;
$n_\chi^{eq}$ is the number
density of the LSP would have in thermal equilibrium at time t;
$H\equiv
\dot{R}/R$ is the Hubble parameter,
$R$ is the cosmic scale factor; $\sigma$
is the cross section for the process
$\chi\chi\rightarrow {\rm ordinary\ matter}$,
$v_{rel}$ is the relative velocity of the two annihilating LSPs, and
the
symbol $\langle\rangle$ denotes a thermal average and
sum over spins of the LSPs. It is clear that rate at which $n_\chi$
decreases is set by two factors. One is the expansion of the
Universe.
The other is due to the excess of the given species annihilation over
the reverse process.

Eq.~(\ref{boltzmanneq})
can be integrated either
numerically or by means of analytic
approximations~\cite{kt,swo,gondolo}.
It occurs that
most methods used in the literature give actually very similar
results (up to a few per cent).
The scaled freeze-out temperature $x_f\equiv T_f/\mchi$ is typically
very small ($x_f={\cal O}(1/20)$), justifying at low temperatures
the approximation
\be
\VEV{\sigma v_{rel}}\simeq a+bx
\label{sigvreleq}
\ee
of the thermally-averaged annihilation cross section
$\VEV{\sigma v_{rel}}$.
The input from high-energy physics is thus included in the
coefficients $a$ and $b$.

Special care must be applied in three cases~\cite{gs}. This is when
the mass of the pair-annihilating species is close to: {\it (i)} half
of
the mass of an {\it s}-channel-exchanged
state (enhanced pole-annihilation); {\it (ii)} the mass of a new
final state (threshold effect); and {\it (iii)} the mass(es) of other
state(s)
with which a DM candidate can also annihilate (co-annihilation
effect).
In such cases the expansion~(\ref{sigvreleq})
fails~\cite{gs}. In particular, the relic density around
poles~\cite{gs,gondolo,nath}
doesn't fall as steeply and is larger than a na{\"\i}ve estimate
would
show.
The co-annihilation effect has been shown to be of great importance
for higgsino-like LSPs and will be discussed in section~\ref{hdm}.

\subsection{LSP Pair-Annihilation Cross Section}\label{pairann}

The DM neutralinos were at the time of `freeze-out' already very
non-relativistic (cold dark matter), and therefore they could
only pair-annihilate into ordinary matter that was {\em less} massive
than
$\mchi$.

The neutralino
relic abundance is thus determined by the annihilation cross section
$\sigma(\chi\chi\rightarrow {\rm ordinary\ matter})$
of the neutralinos into ordinary matter which in turn depends on many
unknown parameters:
$\mu$, $\mone$, $\mtwo$, and $\tanb$, as well as the top quark,
sfermions', and
Higgs masses. In the most general case one needs to specify as many
as 27 parameters, including 21 sfermion masses. The least number of
parameters one has to specify is six: $\mu$, $\mtwo$, and $\tanb$ of
the pure neutralino sector
alone, as well as the top mass $m_t$,
one of the Higgs masses, \eg, the mass of the pseudoscalar $\ma$, and
the mass of the sfermions $\msf$, if assumed degenerate for
simplicity.
With so many free parameters to deal with it is thus {\it a priori}
hard to expect that any solid conclusions can be drawn.
But in fact this is not the case; that is
because, as I will argue later, the LSP relic abundance depends
primarily on only a few parameters, like $\mu$, $\mtwo$ and
the mass of the {\em lightest} sfermion, and only less strongly
on the other parameters. (Similarly, in the context of minimal
supergravity
the number of basic parameters is very limited thus allowing for very
definite predictions; see section~\ref{gutass}.)

Of course, the greater the annihilation rate, the lower the
relic abundance.
The annihilation channels
into ordinary fermion pairs, $\chi\chi\ra\bar f f$, are
always open, except for the top quark which is much heavier.
In addition, the following final states may
become kinematically allowed at larger $\mchi$: Higgs-boson pairs,
vector-boson Higgs boson pairs, and vector-boson pairs.

In the region $\mchi<\mw$ the following annihilation channels
can become at some point kinematically allowed. In addition to
the mentioned channels $\bar f f$, the LSPs can pair-annihilate into
$\hl\hl, \ha\ha,\hl\ha,\hl\hh,\ha\hh,Z\hl$ and $Z\ha$.
For $\mchi>\mw$ new final states open up. They include:
$W^+W^-$, $ZZ$, $C^+C^-$, $HH$, $W^\pm C^\mp$, $ZH$, and $t\bar t$.
It occurs that
in most of the interesting region
of the neutralino parameter space, \ie, for both `mixed' and
gaugino LSPs, it is the annihilation into the Standard Model
fermions that is typically dominant, unless all the sfermion are very
heavy.
The annihilation $\chi\chi\ra\bar f f$ proceeds via the
exchange of the $Z$ and the Higgs bosons in the $s$-channel, and
via the exchange of the sfermions in the $t$- and
$u$-channels. Other final states can, however, also play an
important r\^ole and in a full analysis cannot be neglected.
The dependence of the relic abundance on the parameters
involved is significantly different when the LSP is mostly a
higgsino,
or a gaugino, or else a `mixed' state of both higgsinos and
gauginos. This is of course due to different couplings with which
higgsinos and gauginos couple to other matter.
A closer study shows that the $Z$ exchange is very important
in the almost pure higgsino and `mixed' regions, but not in the
gaugino
region, and the $Z$ pole effect is very broad, unless suppressed by
vanishing coupling to $\chi\chi$ (gaugino and (anti)symmetric
higgsino
$\widetilde H_{(A)S}$). The exchanges of the Higgs bosons are of
significantly lesser importance~\cite{erl}, both because the Higgs
bosons couple only to the `mixed' LSPs and because their couplings to
the ordinary fermions are suppressed by a factor $m_f/\mw$.
Finally, the
effect of the sfermion exchanges is essentially null for higgsino
LSPs, due to a double suppression by a factor $m_f/\mw$ in the
invariant amplitude. On the other hand, sfermion exchange in
gaugino LSP pair-annihilation is dominant, roughly $\abund\propto
\msf^4/\mchi^2$, because the gaugino-fermion-sfermion coupling is not
suppressed.

In figs.~2 and 3 I show a few illustrative
examples of the LSP relic density in the plane $(\mu,\mtwo)$ for a
typical choice
of $\tanb=2$ and $\ma=150$ GeV. In fig.~2 all sfermion
masses are chosen to be $\msf=200$ GeV. (For $\mchi>200$ GeV I take
$\msf=\mchi$.) In fig.~3 I show only the case $\mu<0$
(the case $\mu>0$ is qualitatively similar) and in the left window
all sfermion masses are $\msf=400$ GeV or $\mchi$, whichever is
larger.
In the right window I display the r\^ole of one sfermion: I keep
$\msf=400$ GeV except for one (I choose the selectron) which
is set at $\msel=43$ GeV which is a current lower limit on slepton
masses from LEP.
These figures will be discussed below. As has been already stated in
section~\ref{dmproblem}, $\abund>1$ is cosmologically excluded.
Assuming the theoretically favored value $\Omega=1$ leads to
the approximate
range $0.25\lsim\abund\lsim0.5$. Finally, very small values of
relic abundance (like $\abund\lsim0.025$) are
probably not cosmologically
interesting as they don't solve even the galactic DM problem.

\subsection{Higgsinos as DM}\label{hdm}

Higgsino-like LSPs with mass below
$\mw$, along with gaugino-like
ones (but not the `mixed' states; see next section),
have been considered excellent candidates
for the DM~\cite{enrs,chiasdm}.
In the case of higgsinos, one could easily find
significant regions with the
preferred abundance (except near the $Z$- and Higgs poles) for any
choice of the sfermion masses $\msf$~\cite{erl,enrs,chiasdm} as can
be seen from figs.~2 and 3.
Above $\mw$, the $WW$, $ZZ$, and $t \bar t$ final states cause
the relic abundance of higgsino-like LSPs to drop well below $0.025$,
up to a $\mchi$ of several hundred GeV~\cite{os,gkt}
(see figs.~2 and 3 and, \eg, fig.~2 of the first paper of
Ref.~\cite{os}).
However, as I have mentioned in section~\ref{thcosmlims}, from the
point of view of naturalness higgsinos are not very
attractive~\cite{chiasdm}.

As already mentioned in sect.~\ref{compute},
it has recently been shown that the so-called
`co-annihilation' has a devastating effect on the relic density
of higgsino-like LSPs, making it uninterestingly small. As was
originally pointed out by Griest
and Seckel~\cite{gs}, if there exists another mass state
$\chi^\prime$ with mass only slightly
exceeding the LSP mass, and the cross section for the LSP
pair-annihilation is suppressed relative to the LSP annihilation with
$\chi^\prime$ then it is this latter annihilation that primarily
determines the LSP relic abundance. All the conditions for
co-annihilation are
actually satisfied in the higgsino-like region, where
typically the next-to-lightest neutralino $\chi_2$ and the lightest
chargino $\chi^\pm_1$ are only slightly heavier than the LSP.
In two recent papers~\cite{dn,japan} (see also Ref.~\cite{mcdonald})
it has been estimated that
co-annihilation then greatly
reduces the higgsino-like LSP relic abundance to uninterestingly
small values
($\abund<0.025$). The effect of including co-annihilation on the
figures~2 and 3 is to wipe out all cosmologically
interesting domains from the higgsino regions.
Thus higgsino-like LSPs are neither
theoretically nor cosmologically attractive.
Clearly, finding a higgsino-like LSP at LEP 200, or a chargino that
would correspond to the higgsino region, would thus have dramatic
cosmological implications as it would imply, in the MSSM, a great
deficit of DM.

\subsection{`Mixed' LSPs as DM}\label{mdm}

Next, `mixed' LSPs
(with comparable fractions of both
higgsino and gaugino components)
are probably not cosmologically
attractive either,
as they do not provide enough DM in the Universe~\cite{enrs,chiasdm},
and often even in galactic halos~\cite{chiasdm}.
Typically $\abund$ is found below 0.1, or even 0.025. This is true
both
below and above $\mw$~\cite{chiasdm}
for essentially any choice of independent parameters, like $\tanb$,
$\msf$, or Higgs masses. (See figs.~2 and 3.)
The smallness of $\abund$ results from the
fact that, in the `mixed' region, some of the LSP annihilation
channels always remain unsuppressed up to extremely large values of
the
LSP mass ($\mchi\lsim{\cal O}(100\tev)$~\cite{dn}).

\subsection{Gaugino-like LSP: A Natural Candidate for DM}\label{gdm}

We are left with the gaugino region ($|\mu|\gg\mtwo$).
Here, the crucially important parameters are the sfermion masses.
This is because in this region the sfermion exchange
in the LSP annihilation into $f\bar f$ is dominant:
$\abund\propto \msf^4/\mchi^2$. This can be seen by
comparing figs.~2 and~3.
The $Z$ and Higgs bosons decouple
from gaugino-like neutralinos. For larger $\mchi$ also the final
states
involving Higgs boson pairs become important. This
happens very roughly when {\em all}
the sfermions are heavier than the heavier Higgs bosons (like $\hh$
or $A$).

In the MSSM sfermion masses
are arbitrary. (For simplicity they are often assumed
to be degenerate.) However, as was pointed out in
Ref.~\cite{chiasdm},
it is the mass of the lightest sfermion
(other than the sneutrino, whose
couplings to gaugino-like LSPs are somewhat weaker, and of the stop
whose mass is relevant only for $\mchi>m_t$) that
mostly determines the relic density of the LSPs. In other words:
{\em the gauginos annihilate most effectively via the lightest
charged sfermion exchange}.

The resulting situation can be roughly characterized as follows:
\begin{enumerate}
\item
{\em All the sfermion masses are large, $\msf\gsim 400\gev$.}
(See the left panel of fig.~3; the graph doesn't change
much as one varies $\msf$ from $1\tev$ down to some $400\gev$.)
In this case $\abund>1$ in the region of smaller $\mchi$. At larger
LSP masses, above the kinematic threshold for $\chi$ annihilation
into Higgs-boson final states, the relic abundance drops again,
depending on the actual masses of the Higgs bosons. (The final state
$hh$, which is always open for $\mchi>\mz$, is never
of any significance.)
As one
moves from more pure gaugino-like states towards more `mixed'
states, $\abund$ quickly decreases, and the preferred range
$0.25\lsim\abund\lsim 0.5$ forms only a {\em relatively narrow strip}
in the $(\mu,\mtwo)$ plane.
In the `mixed' region, $\abund$ remains as always below
$0.1$, and often below $0.025$.

Thus, the all-heavy sfermion scenario
is perhaps somewhat less attractive as a solution to the DM
problem~\cite{chiasdm}.

\item
{\em The lightest charged sfermion (say, the selectron $\sel$)
is neither too heavy nor too light,
$120\gev\lsim$
$\msel\lsim400\gev$.}
(Other sfermions could either be also within this mass range, or else
heavy. Their specific values do, however, influence somewhat the
lower bound of $120\gev$: if more sfermions are on the lighter side
then it goes up since more sfermions contribute to reducing
$\abund$.) {\em This is cosmologically the most natural range of
sfermion masses,
for which
large domains of the $(\mu,\mtwo)$ parameter space (for gaugino-like
LSPs) exist with $0.025<\abund<1$}. See fig.~2.
Within those, there are substantial
sub-domains with the preferred abundance $0.25\lsim\abund\lsim0.5$.
Again, the  relic abundance decreases towards `mixed' regions.
Also, it
becomes unacceptably large for asymptotically large values of $|\mu|$
of several TeV.

\item
{\em The lightest charged sfermion is relatively `light',
$43\gev\lsim\msel\lsim120\gev$ (and, of course, $\mchi<\msel,\msf$).}
(Again, other sleptons could be within this mass range, or else
heavier.
As for squarks, one should keep in mind the
current experimental bounds from CDF of $100$--$150$\gev.)
See the right panel of fig.~3.
In this case, the relic abundance in both the gaugino and `mixed'
regions is too small to solve the global DM problem, and often too
small (especially for smaller values of $\msel$) to even
contribute significantly to galactic DM~\cite{chiasdm,chiatlep2}.
If one also includes the effects of co-annihilation in the higgsino
region, one finds no solution satisfying Eq.~(\ref{abundrange}) in
the
whole $(\mu,\mtwo)$ plane. One may thus talk about a {\em
cosmological
lower bound} on $\mchi$~\cite{chiatlep2}, the exact value depending
somewhat on the masses of the heavier sfermions.
\end{enumerate}

It is worth stressing
that, with co-annihilation included, {\em if even a single
slepton were found at LEP (below {\rm43 GeV}) then the LSP relic
abundance
would be $\abund\lsim0.07$ and thus too small to solve the dark
matter
problem.} This can be seen from the right panel of fig.~3.
Similarly, if a slepton were discovered at LEP 200 then the LSP
relic abundance would still be too small~\cite{chiatlep2}
to give $\Omega\simeq1$
for {\em natural} ranges of parameters $\mu$ and $\mtwo$
(roughly below $1\tev$).

Finally,
one might argue that it doesn't matter if the cosmologically favored
region  is `big' or `small': even one point in
the $(\mu,\mtwo)$ plane with large enough relic abundance to solve
the
DM problem (eg.~(\ref{abundrange})) should be enough. This is in
principle true if the minimal supersymmetry model is only considered
without its GUT context. However, when the MSSM is viewed as
resulting
from a more fundamental theory valid at the GUT mass scale, several
other constraints arise, as I will discuss in the next chapter. Thus,
even in the MSSM, solutions which require a certain tuning of
parameters
to solve only the DM problem, are in general less attractive,
although
not excluded. This context should be thus kept in mind in judging
the cases 1 to 3 as being more or less attractive.
What I find important, however, is that, despite a large number of
free parameters, certain general conditions for the
LSP to be the dominant matter component in the flat Universe can
be established and compared with present and future searches for
supersymmetric particles at high energy colliders.

\section{The Effects of Grand Unifications}\label{gutsect}

Until now I have considered the LSP of the minimal supersymmetric
model viewed as a phenomenological theory valid around the Fermi mass
scale.
The MSSM is clearly motivated by grand unifications: one usually
allows for only such soft terms that follow from minimal supergravity
and also assumes that all gauginos share a common (gaugino) mass
$\mhalf$
at the unification scale ($\mone=\mtwo=\mgluino=\mhalf$) which
implies the relations
(\ref{gutone}) and (\ref{guttwo}) at the electroweak scale.
These assumptions themselves are not a part of the definition of
the MSSM (see, \eg, Ref.~\cite{gh}).
They are, however, usually made
in a grand unification scenario, in which the MSSM results from some
underlying GUT structure valid at the scale
$\mx\approx10^{16}\gev$ and is coupled to minimal supergravity.
(See, \eg, the review by Ib\'a\~nez in
Ref.~\cite{susyrev}.)
Below I will consider LSP as DM in  a fuller GUT
framework. Next I will comment on the effect of relaxing GUT
assumptions.

\subsection{Dark Matter in Minimal Supergravity}\label{gutass}

At the GUT scale
all the gauge couplings become equal
${5\over3}\alpha_1=\alpha_2=\alpha_s=\alpha_{GUT}\approx{1\over24}$.
Below $\mx$, the gaugino masses run with the energy scale
in the same way as the
squares of the gauge couplings. In other words, on obtains
$\mone/\alpha_1=\mtwo/\alpha_2=\mgluino/\alpha_s=
\mhalf/\alpha_{GUT}$,
from which the relations (\ref{gutone}) and (\ref{guttwo}) follow, as
well as $\mhalf\approx1.2\mtwo$.

In the absence of a {\em commonly accepted} `standard',
or `minimal' grand
unified theory,
it is usually assumed that, at the GUT scale, the MSSM couples to the
minimal supergravity model which provides
the desired form of the supersymmetry soft breaking terms.
The minimal supergravity model is an effective theory at the GUT
scale.
Its basic
assumptions are described, \eg, in the review by Ib\'a\~nez in
Ref.~\cite{susyrev} (page 56).
I will not review them here but merely quote what is relevant
to the present report.

In addition to assuming a unification of gauge couplings and gaugino
masses,
it is natural to expect that
at the GUT scale also
masses of all the scalars (sfermions and Higgs bosons) are equal to
some common scalar mass $\mnot$.
This assumption,
in conjunction with SUSY and the gauge structure, leads to the
following expressions for the
masses of the sfermions (except for the top squark)
at the electroweak scale (see, \eg, Ref.~\cite{ibanez})
\be
m_{\widetilde\flr}^2 = m_f^2+\mnot^2 + b_{\widetilde\flr}\mhalf^2
\pm\mz^2\cos 2\beta\, [T_3^{\flr} - Q_{\flr}\sin^2\theta_W],
\label{sfmasseq}
\ee
where $\widetilde\flr$ is the left (right) sfermion corresponding to
an
ordinary left (right) fermion, $T_3^{\flr}$ and $Q_{\flr}$ are the
third component of the weak isospin and the electric charge of the
corresponding fermion $f$,
and the coefficients $b$ can be expressed as
functions of the gauge couplings at $\mz$ and are $b\sim6$ for quarks
and $b\sim0.5$ for sleptons.

Thus in the minimal supergravity context there are five fundamental
quantities:
(a) the common gaugino mass $\mhalf$, (b) the common
scalar mass $\mnot$, (c) the Higgs/higgsino mass parameter $\mu$,
(d) the common scale $A\mnot$ of all the trilinear soft SUSY-breaking
terms,
and
(e) the  scale $B\mnot$ of the Higgs soft term
($B\mnot\mu\hone\htwo+\hc$). When the Higgs bosons acquire \vev's,
another independent quantity arises which can be chosen to be $\tanb
=\vtwo/\vone$.
The whole spectrum of masses and
couplings can be parametrized in terms of these six independent
quantities.

It is clear that, in contrast to the MSSM, the masses of the
sfermions and the neutralinos are now linked to each other via
$\mhalf$.
It is thus interesting to see whether the phenomenological and
cosmological relations between the neutralino and sfermion masses,
previously derived at the phenomenological level,
can be accommodated within the minimal supergravity context, and
whether they
provide any constraints on the fundamental supergravity parameters
$\mhalf$ and $\mnot$, and vice versa.
This issue has been studied from various angles
starting with Refs.~\cite{ehn,mihoko}, and more recently in
Refs.~\cite{sugradm,texas,dn,kawasaki} and~\cite{dick}.
The bottom line is that {\em the requirement that the LSP be the
dominant
DM component in the flat ($\Omega=1$) Universe is quite naturally
realized in the context of minimal supergravity but it also imposes
stringent constraints on the fundamental parameters $\mhalf$,
$\mnot$, and $\munot$} (the value of $\mu$ at the GUT scale).
(More specific results are often sensitive to additional
assumptions and simplifications made by various authors.)
For example, it was found in Ref.~\cite{sugradm} (see also
Ref.~\cite{texas}) that the cosmologically
favored region, given by Eq.~(\ref{abundrange}), is only (except near
poles) realized in
the narrow strip of comparable values of $\mnot$ and
$\mhalf\simeq1.2\mtwo$ (see fig.~4). The region $\mnot\gg\mhalf$ was
found cosmologically excluded ($\abund>1$). In the region
$\mnot\ll\mhalf$ there was always very little DM
($\abund<0.025$). Cosmologically (\ie, requiring the LSP to be the
dominant mass component of the flat Universe)
one can derive a lower bound on $\mnot$:  $\mnot\gsim100\gev$ for
$\mchi<\mw$cite{sugradm}, and $\mnot\gsim40\gev$
in general~\cite{dn}.
This, along with other (experimental and theoretical)
constraints, allows for rather specific predictions at the
electroweak
energy scale. For example, one finds a lower bound on the mass of
the lightest slepton of about $100\gev$~\cite{dn} (see also
Ref.~\cite{texas-new}),
thus confirming a result of Ref.~\cite{chiatlep2}. As for the
squarks,
which in minimal supergravity are invariably heavier, one {\em
typically} gets $m_{\widetilde q}$ significantly exceeding
200\gev~\cite{texas-new,dick}, although Drees and Nojiri~\cite{dn} in
certain less favored cases (rather large values of  $A\mnot$) find
$m_{\widetilde q}$ even below 200\gev.

Roberts and Ross~\cite{rr} have recently performed a careful
study of the phenomenological implications of the grand unifications
with the electroweak gauge symmetry triggered by supersymmetry
breaking terms.
In their analysis, they took into account the corrections to the
running
of the various parameters of the model (gauge couplings and masses)
due to multiple mass thresholds above the electroweak energy scale.
The underlying idea is that above the SUSY breaking scale $\msusy$
one uses fully supersymmetric renormalization group equations (RGEs)
to evolve the model's parameters from the GUT scale down. However, as
the masses of the heaviest supersymmetric particles (typically the
gluino and squarks) become larger than the energy scale $Q$ at which
they are evaluated, one needs to decouple them from the running of
the RGEs. This has to be successively done with every new state
(mass-threshold), as
one further evolves down to the $\mz$ scale.
These thresholds were shown to be of significant importance.
Another remarkable feature of their analysis was that
that they did not fix any of the constraints at the start (like often
made $m_{\widetilde q}<1\tev$) but instead compared
their relative effects on the final output.
On the basis of several independent criteria (bounds on $m_t$ and
$m_b$, improved naturalness
criterion, and improved experimental bounds on $\alpha_s$ and other
electroweak and supersymmetric quantities),
Roberts and Ross generally found both $\mhalf$ and
$\mnot$ to lie within several TeV, with smaller values (a few hundred
GeV) generally favored by the values of $\alpha_s$ and the
naturalness
criterion.
The allowed parameter space, however, still
remained quite large.

Their analysis was next extended~\cite{dick} to include
the requirement that the LSP be the dominant dark matter
component of the flat Universe, see figs.~5
and~6.
This constraint was found to be of particular importance.
First, in the parameter space not experimentally excluded by LEP and
the Tevatron the lightest neutralino was found to be always
the LSP. Furthermore, it invariably came out to be gaugino-like,
(see fig.~5d)
in perfect agreement with previous
conclusions~\cite{chiasdm,dn,japan}.

Remarkably, the dark matter
constraint was found to be consistent with other bounds: the favored
range of the
bottom quark mass ($4.15\gev\lsim m_b\lsim4.35\gev$~\cite{bottom});
the expected range
of the top mass ({\em implying} a rather large value
$m_t\gsim150\gev$); and the present experimental bounds on the strong
coupling $\alpha_s(\mz)=0.122\pm0.010$~\cite{altar}.
All these criteria can be
simultaneously satisfied (fig.~6) without any excessive
fine-tuning of
parameters. Furthermore, the
dark matter constraint eliminated large fractions of the parameter
space corresponding to $\mnot\gg\mhalf$ (fig.~5c). Also
the CDF limit $m_t>91\gev$ eliminated the region
$\mnot\ll\mhalf$ where the LSP is invariably higgsino-like for which
the relic abundance has been found to be very small.

The region of
the parameter space consistent with all the above constraints is
severely limited but, remarkably, consistent with
supersymmetric masses below 1~TeV.
In particular, imposing a requirement that the LSP provide most DM
in the flat Universe again provides
very stringent cosmological limits~\cite{chiatlep2} on the masses of
the LSP and other particles. One also finds~\cite{dick}
\be
60 ~{\rm GeV}\lsim\mchi\lsim 200 ~{\rm GeV},
\label{lsprange}
\ee the upper limit being also expected
in the
minimal supersymmetric model~\cite{chiasdm,bg} on the basis
of naturalness. Similarly, one obtains~\cite{dick}
\bea
150 ~{\rm GeV}\lsim\mcharone\lsim300~{\rm GeV}\\
200 ~{\rm GeV}\lsim\msl\lsim500 ~{\rm GeV}\\
250 ~{\rm GeV}\lsim\msq\lsim850 ~{\rm GeV}\\
350 ~{\rm GeV}\lsim\mgluino\lsim900 ~{\rm GeV}.
\label{massranges}
\eea
The heavy Higgs bosons are roughly in the mass range between 250 GeV
and 700 GeV. Of course, lower values of all these masses correspond
to less fine tuning and larger values of $\alpha_s$.
The lightest Higgs boson tree-level mass invariably
comes out close to $\mz$; its one-loop-corrected
value~\cite{radcorrs} is then roughly in the range 120 to 150 GeV.
As a bonus, one also finds~\cite{dick}
$0.116\lsim\alpha_s(\mz)\lsim 0.120$.

In summary,  the
spectrum of supersymmetric particles is expected to lie well within
the discovery
potential of the LHC and the SSC. However, if
the LSP is to be the dominant component of matter in the flat
Universe,
then it perhaps less likely, although not impossible,
for SUSY to be discovered at LEP 200 and/or the Tevatron.

Of course, the specific numerical results of this and other analyses
should be taken with a grain of salt. It suffices to say that
the findings of recent studies~\cite{sugradm,dn,texas,texas-new}
are in general consistent with each other. More detailed
comparisons are, however, almost impossible due to somewhat
different assumptions and approximations used by various authors.
Also, additional effects
(like radiative corrections to Higgs masses, or mass-thresholds
around the GUT scale) and complications to
the model may alter them somewhat. Nevertheless, it is rather
encouraging that the `minimal' set of assumptions along with
a number of {\it a priori} independent criteria provide
self-consistent
and reasonable predictions for minimal supersymmetry coupled to
minimal supergravity.

\subsection{Effect of Relaxing GUT Assumptions}\label{griest}

Finally, one shouldn't forget that the concept
of grand unification (and the related assumptions), while attractive
and in some sense natural, will most likely
not be directly tested experimentally in the
foreseeable future (perhaps never), even though recent LEP results
can be
interpreted as supporting it.
Moreover, recent results in superstring theory indicate that the
usual GUT assumptions about the common scalar and gaugino masses need
not be necessarily hold~\cite{munoz}.
One would therefore like to learn to what
extent the various phenomenological and cosmological results are
modified when the simple GUT relations given above
are relaxed.
This question was addressed in Ref.~\cite{nogut} (see also
Ref.~\cite{nogutjapan})
in the context of the minimal supersymmetric model {\em without} the
GUT assumptions about the gaugino masses $\mone$, $\mtwo$, and
$\mgluino$ (nor of course the supergravity relations between scalar
masses).
Relaxing GUT assumptions results in potentially significant
modifications of the phenomenology and cosmology of the
neutralino LSP. The study~\cite{nogut} focused on a light neutralino
LSP, within
the mass range of a few tens of\gev, since such neutralinos will be
searched for in accelerators and astrophysical experiments during the
next decade.
The results can be briefly summarized as
follows:
\begin{enumerate}
\item
As expected, the experimental lower bound $\mchi\gsim20\gev$
doesn't exist anymore,
since the gluino mass bound from CDF no longer applies;
in fact no experimental bound can be derived in general.
(This is not surprising: it is not always remembered that while
experimental constraints on charged supersymmetric particles can be
regarded as relatively general, the corresponding bounds on the
neutral
particles (like the Higgs bosons or the neutralinos) are very
model-dependent, or even assumption-dependent. In a sense,
experimental
lower bounds on neutral particles could be regarded as (upper)
limits of potential experimental ability to search for neutral states
{\em in a given model}.)
However, even though experimentally very light neutralinos are now
again allowed, it is well known that
cosmologically neutralinos below a few\gev\ are forbidden,
as they would overclose the Universe (the Zel'dovich bound).
\item
The indicative theoretical {\em upper} bound $\mchi\lsim150\gev$
does not apply either for the same reason as above. In principle the
LSP
could be very heavy, although perhaps one should expect
$\mchi\lsim1\tev$ for naturalness reasons.
\item
The composition and mass contours also become significantly
modified, depending on how different the ratio $r\equiv\mone/\mtwo$
is
from the GUT case ($\rgut\approx 0.5$). Since in the gaugino
region $\mchi\approx\mone$, one finds an interesting scaling
property~\cite{nogut}
in the ($\mu,\mtwo$) plane: in the experimentally allowed region, for
smaller values of $r$ ($r<\rgut$), generally lighter neutralinos are
allowed, whilst for $r>\rgut$ gaugino-like neutralinos are generally
heavier than in the GUT case. The higgsino region as a whole is
also shifted to larger (smaller) values of $\mtwo$ for smaller
(larger)
values of $r$, but otherwise the mass composition contours are not
greatly modified.

\item
Cosmologically allowed/favored/excluded regions are also
significantly affected by the scaling property,
depending on the choice of $r$. For $r\ll 1$, the gaugino-like LSP
easily gives
the closure density. (In contrast, when $r\gg 1$, the LSP is often
predominantly wino-like and, due to co-annihilation effects, its
relic
density is very small~\cite{nogutjapan}).
\end{enumerate}

In conclusion, even modest modification of the usual grand
unification assumptions has a considerable impact on the
phenomenological and cosmological properties of the LSP. In
particular,
the gaugino-like LSPs as light as a few\gev\ could
still be interesting candidates for
DM. Experimentalists in their search for new physics should not be
constrained by theoretical expectations and biases.

\section{Direct and Indirect Searches for Supersymmetric Dark Matter}
\label{dmsearches}

I will only briefly outline the present status and future prospects
for
the neutralino DM searches. More details can be found, \eg, in
Ref.~\cite{caldwell}. The underlying assumption is that the LSP is
a dominant component of the halo of our Milky Way, with the relic
density of $\rho_\chi\simeq0.3\gev/{\rm cm^3}$  (about 3000 LSPs with
mass $\mchi=100\gev$ per cubic meter) and the mean velocity
of about 300~km/s.

One can search for DM neutralinos either {\em directly}, through the
halo LSP elastic scattering off nuclei $\chi N\ra \chi N$, or {\em
indirectly}, by looking for traces of decays of LSP pair-annihilation
products (mainly neutrinos).

At present direct searches rely on `warm' germanium and silicon
detectors and lack some two order of sensitivity necessary to probe
the neutralino DM. The problem lies in the fact that neutralinos,
being Majorana particles, couple to nuclei proportional to their spin
(although there also exist some coherent interactions with nuclei
without spin), and the related scattering cross sections are
invariably
very small.
Significant progress
is reportedly~\cite{caldwell,sadoulet} being made in the R\&D of
cryogenic (T$<$0.03\,K) detectors. It is expected that
they will achieve desired sensitivity by a dramatic noise reduction
(mostly $\beta$-radiation and Compton electrons) by measuring the
energy
deposition (a few tens of \kev) in both the phonons from nuclear
recoil and ionization. First working modules should be ready by
Spring'93 and a full multi-kg detector by  the Fall of 1994. There
are
also plans to replace the presently used spinless $^{76}$Ge with
with large-spin nuclei (like $^{73}$Ge with spin=9/2). In Europe,
another promising and innovative technique based on superconducting
detectors is currently being developed. (More details can be
found, \eg, in Ref.~\cite{caldwell}.)

Indirect searches have been looking for high-energy neutrinos and
up-going muons in underground detectors. The basic idea can be
briefly
outlined as follows. Halo LSPs can be captured in celestial bodies
(Sun, Earth) through their scattering off heavy (core)
nuclei. If the scattering is sufficiently frequent, they loose enough
energy to become gravitationally trapped. After enough LSPs have
accumulated, they can again start pair-annihilating into leptons,
quarks, and gauge and Higgs bosons, most of which immediately decay
into lighter states. Only neutrinos can make their way out of the Sun
or Earth core. Their energy is roughly
$E_\nu\sim({1\over3}-{1\over2})\mchi$. They can pass through
underground neutrino detectors or, by hitting a nearby rock, produce
secondary up-going muons that can be
searched for in underground muon detectors. Estimates show that for
$\mchi\gsim80\gev$ signal from the Sun should be dominant. Lighter
neutralinos could more effectively scatter off  nuclei in the Earth's
core (which are of comparable mass) and thus produce a stronger
signal.

It is a bit unfortunate that the LSPs that most effectively interact
with nuclei are of the `mixed' type which typically give very
little dark matter and thus are not the best candidates for the DM.
(Their pair-annihilation is in that case also very efficient, see
section~\ref{mdm}.) Nevertheless,
some regions of the LSP parameter space have already been
explored by the IMB and Fr{\'e}jus groups, and more recently by
Kamiokande~\cite{kamioka} in searches for up-going muons.
Whether or not those regions have been excluded depends on what
fraction of Galactic halo is shared by the neutralinos.

Neutrino detectors (like, \eg, MACRO, DUMAND, AMANDA),
and a new generation of cryogenic detectors should have a much larger
potential to fully cover the neutralino parameter space. Much work
has, and is being, done both on a technical
front and in improving existing theoretical calculations of
scattering
cross sections, capture
rates, \etc~\cite{searchprogress}. First decisive results are
expected within the next few years.

A whole spectrum of other methods have been considered. They include
looking for products of LSP pair-annihilation in the
Galactic halo, like monochromatic photons, $e^{+}$-line radiation,
and
in continuum spectrum of cosmic antiproton, positron, and
gamma--rays, to mention just of few. All these techniques seem to
suffer from large theoretical and observational uncertainties and
insurmountable backgrounds. At
present they are considered less promising.

\section{Agnostic Comments}
\label{agnostic}

Before we all (I mean supersymmetry enthusiasts) become overly
excited
I would like to remind that:
\begin{itemize}
\item
Supersymmetry has not been discovered (yet). No `hints' nor
`evidence', while encouraging, cannot be a substitute for
a truly unambiguous experimental signal.
\item
Dark matter has also not been discovered yet. Same remarks apply.
Moreover, even its nature (macroscopic objects vs. particles,
baryons vs. weakly interacting species, hot vs. cold DM) and
abundance haven't been definitely established.
\item
Cosmic inflation and $\Omega=1$ still remain (very attractive)
hypotheses for which there is at best some observational support.
\item
The LSP may after all not be completely stable. The $R$-parity,
invoked to insure the stability of the proton, is not a `fundamental'
symmetry, and can easily be relaxed. For example, one can allow for
lepton or baryon number violating interactions (but not both), in
which
case the LSP would be unstable. Despite stringent limits, such a
scenario may be easily realized and in fact may be favored in light
of the COBE discovery.
\item
Simple (simplistic?) assumptions at the GUT scale can easily be
modified or relaxed, even in the context of superstrings, and thus
may lead to significantly altered conclusions at the electroweak
scale.
\item
Minimal SUSY may not be the end of the story! After all, it
is not favored by any particular reason other than simplicity.
On the other hand, it is encouraging that even the simplest
supersymmetric model is both theoretically and cosmologically
attractive, while not running into any  immediate conflict with
experiment.
\item
Finally, one may argue that still lacking is a satisfactory
explanation of why the LSP relic abundance should be close to unity
in
the first place. (See,
however, Ref.~\cite{turnerinsm}.) Why/how did supersymmetric
parameters
`conspire' to make the LSP dominate the Universe? This question
should
be answered (or maybe shown ill-posed?) by a truly fundamental
theory.
\end{itemize}

\section{Final Comments}
\label{summary}

In this review I have addressed the issue of whether supersymmetry
can
provide an attractive candidate for solving the dark matter problem.
I
have argued that the lightest neutralino of the minimal supersymmetry
has all the desired properties for being both the LSP and DM.
This is certainly encouraging given the fact that supersymmetry is
the
leading candidate for the extension of the Standard Model. In
addition,
the experimental values of the gauge coupling, as measured at LEP,
when
evaluated in the minimal supersymmetric model,
become equal at the SUSY GUT scale (of around $10^{16}\gev$), thus
supporting the idea  of supersymmetric unification.

Cosmology provides additional constraints on the supersymmetric
parameter space. Requiring that the LSP provide enough
(or at least a substantial fraction of) DM in the flat
Universe, when combined with the usual theoretical assumptions and
present experimental bounds, points towards the gaugino
in the mass range of  several tens of GeV as the most natural
candidate
for both the LSP and the DM. Significant fractions
of this mass range should be accessible to LEP 200
and planned experiments for DM searches.
In addition, cosmologically favored masses for sleptons and squarks
lie
beyond the reach of LEP 200 and the Tevatron (what may be somewhat
discouraging) but well within the reach of the future
hadronic supercolliders: the LHC and the SSC.
During the next decade, accelerator and astrophysical DM searches
should be able to test this and other predictions of supersymmetry.

It is not unlikely that
most of the Universe is actually made of supersymmetric particles
while what we usually call `ordinary' matter may look more like a
very
exotic component of the cosmic zoo. (One might look at this as yet
another step in the Copernican revolution!) Nothing short of a direct
SUSY discovery will convince the whole physics community but we
should certainly be encouraged to continue vigorous searches. It is
not unlikely that the first signal of supersymmetry may actually come
from searches for dark matter.

\vskip 1cm
\section*{Acknowledgments}

\noindent
I would like to thank Professor A. Bia{\l}as, the Director of the
Ettore Majorana Centre, Professor A.~Zichichi,
the organizers of the Erice Workshop, Dr. L.~Cifarelli and Professor
V.~Khoze, and Professor H.~Haber for their kind invitation to the
respective meetings.

\bigskip
%		\newpage
%%%%%%%%%%%%% begin refs %%%%%%%%%%%%%%%%

%
\newpage
\section*{Figure Captions}
\vskip 0.1cm
%	\listoffigures

\noindent{\bf Figure 1}: Contours of the neutralino mass and
gaugino-higgsino
compositions (purity, as defined in the text)
in the ($\mu,\mtwo$) plane for $\tanb=2$ and
for $\mu<0$ (left panel) and $\mu>0$ (right panel).
Lightest neutralino mass contours
are labeled (in GeV).  Contours of constant gaugino purity
($p_{gaugino}=Z_{11}^2 + Z_{12}^2$) are shown with $p_{gaugino}=$
0.99, 0.9, 0.5, 0.1, and 0.01 from larger $|\mu|$ to
larger $\mtwo$.  (Note
$p_{gaugino} = 0.01$ implies a 99\% higgsino.)
In the regions labeled $\widetilde B$ and  limited by the dashed
curves
$p_{\widetilde B} \geq 0.99$ (almost pure bino), while the
region (labeled $\widetilde H_S$) above the other
dashed curve correspond to almost pure symmetric
higgsino ($p_{higgsino} > 0.99$). (The almost pure anti-symmetric
higgsino $H_A$ lies above the range of $\mtwo$ in the window
$\mu>0$.)
The areas marked ``LEP" are ruled out by LEP
and the areas below the curves
marked ``CDF" are ruled out by CDF.
Also marked is the remaining photino region with 95\% purity. \\
\noindent{\bf Figure 2}: Contours of relic abundance in the
($\mu,\mtwo$)
plane, with $\tanb=2$, $\ma=150\gev$,
$\msf=200$ GeV  or $\mchi$ (whichever is larger) for all sfermions,
and $\mu<0$ ($\mu>0$) in the left (right) panel.
Areas excluded by LEP (CDF) are marked ``LEP'' (``CDF''). The
relic density increases with increasing grayness. The cosmologically
excluded region $\abund>1$ is delineated by
a thick solid line in the grey region. The cosmologically
favored regions ($0.25\lsim\abund\lsim0.5$) are delineated by dotted
lines. The long-dashed and
thick-solid contours correspond to 0.1 and 0.025, respectively.
The LSP co-annihilation has not been included. Its effect
would be to greatly reduce the relic abundance in the higgsino region
($\mtwo\gg|\mu|$). \\
\noindent{\bf Figure 3}: Contours of relic abundance in the
($\mu,\mtwo$)
plane, with $\tanb=2$, $\ma=150\gev$, and $\mu<0$. (The case $\mu>0$
is qualitatively similar.) In the left panel all $\msf=400$ GeV  or
$\mchi$ (whichever is larger). In the right panel all sfermion
masses are as before but for the selectron $\msel=45$ GeV or
$\mchi$ (whichever is larger). All the textures are as in fig.~2.
The LSP co-annihilation has not been included. Its effect
would be to greatly reduce the relic abundance in the higgsino region
($\mtwo\gg|\mu|$). \\
\noindent{\bf Figure 4}: Contours of relic LSP density in the
$(\mu,\mtwo)$
plane for $\tanb=2$, $\ma=200\gev$, and $\mu=\mp500\gev$ in the
left (right) panel. The cosmologically favored regions with
$0.25 < \abund < 0.5$ are shaded, while regions where
$\abund > 1$ are cross-hatched, and regions excluded by
LEP and CDF are hatched. The $\abund = 0.1$ contour
is shown as a thin dashed line.
Above the lines
``1" $m_{\tilde q}>$ 1 TeV for at least one squark,
above ``2" $m_{\tilde g}>$ 1 TeV, and to the left of ``3"
$m_{\tilde q}<$ 100 GeV for at least one squark. \\
\noindent{\bf Figure 5}: In the plane ($\mhalf,\mnot$) for the fixed
ratio
$\munot/\mnot=2$ I show: in window a) the mass contours of the top
and the bottom quarks (solid and short-dashed lines, respectively);
in window b) the contours of $\alpha_s(\mz)$ (solid) and the measure
$c$ of fine-tuning (dots); in window
c) the relic
abundance $\abund$ of the LSP; and in window d) the mass contours of
the LSP (solid) and the lightest chargino (dashed) at 50, 100, 150,
200, 500, and 1000 GeV, starting from left,
and the contribution (dots)
of the
bino to the LSP composition (bino purity).
In all the windows thick
solid lines delineate regions experimentally excluded by the CDF
(marked CDF) where $\mtop<91$ GeV and by the LEP experiments (LEP)
where the lightest chargino is lighter than 46 GeV. In window c) we
also mark by $\abund>1$ the region cosmologically excluded (too young
Universe). The thin band between the thick dashed lines in window c)
corresponds to the flat Universe ($\Omega=1$). In window d) the
region excluded by CDF almost coincides with
the bino purity of 50$\%$ or less. \\
\noindent{\bf Figure 6}: I show a blow-up of the down-left portion of
the plane
($\mhalf,\mnot$) from the previous figure for the same fixed ratio
$\munot/\mnot=2$.  I combine the mass contours of the top and the
bottom quarks with the ones  of the LSP relic mass density. I use
the
same textures as in Fig.~5 but I also show
(two medium-thick short-dashed lines) the contours $\mbot=4.15$ GeV
and
4.35 GeV which reflect the currently favoured range of the mass of
the
bottom quark (see text). I see that they cross the cosmologically
favored region (thick long-dashed lines) marked $\Omega=1$ at roughly
150 GeV$\lsim\mhalf,\, \mnot\lsim$ 400 GeV and for $\mtop$ broadly
between 150 GeV and 180 GeV.
%
%%%%%%%%%%%%% end of refs %%%%%%%%%%%%%%%
\end{document}